\documentclass[%
 amsmath,amssymb,
 aps,
]{revtex4-2}

\usepackage{soul} % for the command \hl
\usepackage{graphicx}% Include figure files
\usepackage{dcolumn}% Align table columns on decimal point
\usepackage{bm}% bold math
\usepackage{hyperref}% add hypertext capabilities
\usepackage{mathtools}% Loads amsmath
\usepackage{xcolor}
\begin{document}

\title{Darkness cannot bind them:\\ a no-bound theorem for $d=5$ Myers-Perry null \& timelike geodesics}%

\author{João P. A. Novo}
\affiliation{Departamento de Matem\'atica  da Universidade de Aveiro and
	Center for Research and Development in Mathematics and Applications (CIDMA),
	Campus de Santiago, 3810-193 Aveiro, Portugal}
\email{j.novo@ua.pt}

\date{\today}% It is always \today, today,

\begin{abstract}
In Newtonian gravity, it is well known that Kepler's problem admits no bound solutions in more than three spatial dimensions. This limitation extends naturally to General Relativity, where Tangherlini demonstrated that Schwarzschild black holes in higher dimensions admit no bound timelike geodesics. However, an analogous result for the rotating counterpart of the five-dimensional Tangherlini spacetime—the $d=5$ Myers-Perry black hole—has not yet been established. This work addresses this gap by proving that no bound timelike geodesics exist outside the event horizon of a $d=5$ Myers-Perry black hole, for any choice of spin parameters that avoid naked singularities. With this result in place, we further generalize to null geodesics. It is shown that radially bound null geodesics, which are absent in the four-dimensional Kerr spacetime as established by Wilkins, also cannot exist in the $d=5$ Myers-Perry spacetime. These results complete the geodesic analysis of this spacetime and provide a direct generalization of Wilkins' classical result to higher dimensions. Specifically, we establish the following theorem: no radially bound timelike or null geodesics are possible outside the event horizon of a $d=5$ Myers-Perry black hole, regardless of the spin configuration.
\end{abstract}

%Use showkeys class option if keyword
\maketitle

\tableofcontents

\section{\label{sec:Int} Introduction}
%%%%%%%%%%%%%%%%%%%%%%%%%%%%%%%%%%%%%%%%%%%%%%%%%%%%%%%%%%%%%%%%%%%%%%%%%
%%%%%%%%%%%%%%%%%%%%%%%%%%%%%%%%%%%%%%%%%%%%%%%%%%%%%%%%%%%%%%%%%%%%%%%%%

In 1972, Wilkins studied the geodesics in the Kerr spacetime~\cite{Wilkins:1972rs}, establishing the conditions for radially bound timelike geodesics, and proving that null geodesics cannot be radially bound outside the event horizon. Therefore, although massive test particles can be trapped inside a range of the radial coordinate, photons (or other null particles) are either deflected to infinity or absorbed by the black hole. As a consequence, there are no stable photon orbits in the Kerr solution. In this context, a \emph{bound} trajectory is defined as one confined to a \emph{finite} range of radii. This definition excludes circular orbits, in particular timelike circular orbits (TCOs) and light rings (LRs), nonetheless it should be remarked that the existence of stable circular orbits implies the existence of bound trajectories. The goal of this paper is to perform a similar analysis considering the higher dimensional generalization of the Kerr black hole, the $d=5$ Myers-Perry (MP) black hole.

Radially bounded motion in $3$ spatial dimensions is possible due to a balance between the attractive gravitational field, which decays as $1/r$, and a repulsive centrifugal force due to angular momentum, which decays as $1/r^2$. While the decay of the centrifugal force is independent of the spatial dimensionality, the same is not true for the gravitational field, which,in $n$ spatial dimensions, decays as $1/r^{n-2}$. This dependence on the dimensionality implies that there are no stable bound orbits for the Keplerian problem for $n>3$ spatial dimensions. So is the case for the hyperspherically symmetric Tangherlini solution~\cite{Tangherlini:1963bw}. This matching of decays of attractive and repulsive forces allows for black hole solutions with non topologically spherical black hole horizons~\cite{Emparan:2001wn,Emparan:2008eg}.

The extension of this result to the rotating version of the $d=5$ Tangherlini black hole, the MP solution, has been considered several times in the literature before, although restricted to specific geodesically complete hypersurfaces, usually dubbed the  ``equatorial planes" of the $d=5$ MP solution \cite{Frolov:2003en,Cardoso:2008bp,Dadhich:2020svz,Dadhich:2021vdd}, as an analogy with the Kerr solution. However, this terminology is misleading. These hypersurfaces correspond to the fixed points of the rotational Killing vector fields where the associated angular momentum (Noether charge) must vanish, and thus are more akin to the rotation axis of the Kerr metric. Therefore the study of these hypersurfaces does not reveal the complete story of timelike geodesics in the $d=5$ MP solution, for comparison it is also impossible to have bound geodesics restricted to Kerr's rotation axis. Other attempts have been made that consider limiting cases of the $d=5$ MP solution. This paper will extend these partial results to the most general case, that is any values of the angular momenta and any values of the spin parameters (as long as they prevent naked singularities), definitively asserting that the $d=5$ MP spacetime cannot harbour radially bound timelike geodesics outside the event horizon.

It is possible to study the null geodesics of the spacetime by taking the null limit of the aforementioned analysis. The null geodesic flow near a black hole is directly connected to observational signatures both in the electromagnetic and gravitational waves channel. In the former, the shadow of a black hole is determined by a special set of unstable photon orbits~\cite{Bardeen:1973tla,Falcke:1999pj}, in the latter, the early-time ringdown signal of a perturbed black hole is related to the properties of the unstable light rings~\cite{Goebel,Cardoso:2016rao}. It is also conjectured that the existence of radially bound null geodesics can lead to a build up of energy and trigger non-linear instabilities~\cite{Cardoso:2014sna,Keir:2014oka} which have been observed in some models~\cite{Cunha:2022gde}. Similarly to the timelike geodesics, it is shown that the $d=5$ MP spacetime cannot harbour such stable null orbits.

Recently, the concept of \emph{hypershadow} appeared in the literature as the volumetric shadow of some higher dimensional black object~\cite{Novo:2024wyn}. Where the $d=5$ cohomogeneity-one MP solution was analyzed. It was shown that, as in the Kerr spacetime, there are no radially bound null geodesics outside the event horizon, meaning that the (hyper)spherical photon orbits determine the edge of the hypershadow. Therefore, this theorem is here extended to the most general case, implying that the hypershadow of a general MP black hole, free of naked singularities, is also determined by its spherical photon orbits.

This paper is organised as follows. In Sec. \ref{sec:MP_Solution} the $d=5$ MP solution is presented and briefly discussed, and the equations of motion are presented. In Sec. \ref{sec:The_Theorem_Time} it is established that there are no radially bound timelike geodesics in the general $d=5$ MP solution. This result is extended to the null case in Sec. \ref{sec:The_Theorem}, the particular case of the extremal single rotating solution is discussed in detail. The conclusions are presented in Sec. \ref{sec:Conclusions}.

%%%%%%%%%%%%%%%%%%%%%%%%%%%%%%%%%%%%%%%%%%%%%%%%%%%%%%%%%%%%%%%%%%%%%%%%%
%%%%%%%%%%%%%%%%%%%%%%%%%%%%%%%%%%%%%%%%%%%%%%%%%%%%%%%%%%%%%%%%%%%%%%%%%
\section{The Myers-Perry Solution\label{sec:MP_Solution}}
%%%%%%%%%%%%%%%%%%%%%%%%%%%%%%%%%%%%%%%%%%%%%%%%%%%%%%%%%%%%%%%%%%%%%%%%%
%%%%%%%%%%%%%%%%%%%%%%%%%%%%%%%%%%%%%%%%%%%%%%%%%%%%%%%%%%%%%%%%%%%%%%%%%

Put forth in 1986 the Myers-Perry solutions \cite{Myers:1986un,Myers:2011yc}
describe topologically spherical black holes for arbitrary $d>4$
dimensions, such solutions rotate in $\left[(d-1)/2\right]$ orthogonal planes,
giving rise to the same number of independent angular momenta. These
generalise the Kerr solution to higher dimensions.
For $d=5$ the solution takes the form (here the coordinates of ~\cite{Frolov:2003en} are used):

\begin{align}
\mathrm{d}s^{2}= & -\mathrm{d}t^{2}+\left(x+a^{2}\right)\sin^{2}\theta\mathrm{d}\phi^{2}+\left(x+b^{2}\right)\cos^{2}\theta\mathrm{d}\psi^{2}\nonumber \\
 & \quad+\frac{\mu^{2}}{\rho^{2}}\left(\mathrm{d}t+a\sin^{2}\theta\mathrm{d}\phi+b\cos^{2}\theta\mathrm{d}\psi\right)^{2}+\frac{\rho^{2}}{4\Delta}\mathrm{d}x^{2}+\rho^{2}\mathrm{d}\theta^{2}\,,\label{eq:MP_line}
\end{align}
with
\begin{equation}
\rho^{2}=x+a^{2}\cos^{2}\theta+b^{2}\sin^{2}\theta\,,\quad\Delta=\left(x+a^{2}\right)\left(x+b^{2}\right)-\mu^{2}x\,.
\end{equation}
The Arnowitt–Deser–Misner (ADM) mass of this spacetime is $M=3\pi\mu^{2}/(8G)$, and the two
independent angular momenta are proportional to the spin parameters
$a$ and $b$, and given by
\begin{equation}
J_{a}=\frac{2}{3}Ma\,,\quad J_{b}=\frac{2}{3}Mb\,.
\end{equation}
To simplify some expressions the radial coordinate is not the usual
Boyer-Lindquist type coordinate, $r$, instead the metric is written
in terms of $x=r^{2}$ \cite{Frolov:2003en}. At spatial
infinity the metric reduces to flat spacetime in Hopf coordinates,
the ranges of the angular coordinates are $\theta\in\left[0,\pi/2\right]\,,\phi\in\left[0,2\pi\right[\,,\psi\in\left[0,2\pi\right[$.
The metric possesses two horizons at the roots of $\Delta$, with
the outermost root being
\begin{equation}
x_{H}=\frac{\mu^{2}-a^{2}-b^{2}+\sqrt{\left(\mu^{2}-a^{2}-b^{2}\right)^{2}-4a^{2}b^{2}}}{2}\,.
\end{equation}
To avoid a naked singularity the spins must satisfy by $\mu\geq\left|a\right|+\left|b\right|$.
Since this work will be focused only on the region outside the event
horizon the radial coordinate will be constrained to $x\geq x_{H}$.
The event horizon is the Killing horizon of the Killing vector field
$\chi=\partial_{t}-\Omega_{a}\partial_{\phi}-\Omega_{b}\partial_{\psi}$,
from where it is possible to read the horizon's angular velocities
$\Omega_{i}=i/\left(x_{H}+i^{2}\right)\,,i=a,b$.

\subsection{Equations of motion}

The $d=5$ Myers-Perry spacetime is completely integrable~\cite{Vasudevan:2004mr}, as besides the Noether charges $p_{t}=-E\,,p_{\phi}=\Phi$
and $p_{\psi}=\Psi$ associated with the three Killing vector fields
$\partial_{t}\,,\partial_{\phi}$ and $\partial_{\psi}$, the metric possesses a non-trivial Killing tensor. This Killing tensor allows for a complete separation of the Hamilton-Jacobi equation, due to the existence  a Carter-like separation constant~\cite{Carter:1968rr,Frolov:2006dqt}, denoted $\mathcal{Q}$. 
For a test particle of mass $m$ the Hamilton-Jacobi equation is $-m^2=g^{\mu\nu}p_\mu p_\nu$, which, after performing the substitutions reads
\begin{align}
    -m^{2}\rho^{2}=&p_{\theta}^{2}+\frac{\Phi^{2}}{\sin^{2}\theta}+\frac{\Psi^{2}}{\cos^{2}\theta}-E^{2}\left(a^{2}\cos^{2}\theta+b^{2}\sin^{2}\theta\right)+4p_{x}^{2}\Delta-xE^{2}\\&\quad-\mu^{2}\frac{\left(x+a^{2}\right)\left(x+b^{2}\right)}{\Delta}\left(E+\frac{a}{x+a^{2}}\Phi+\frac{b}{x+b^{2}}\Psi\right)^{2}\\&\quad-\left(a^{2}-b^{2}\right)\left(\frac{\Phi^{2}}{x+a^{2}}-\frac{\Psi^{2}}{x+b^{2}}\right)\,,
\end{align}
After some rearrangements this equation can be written as $\mathcal{H}_x(x)+\mathcal{H}_\theta(\theta)=0$, with
\begin{align}
    \mathcal{H}_x&=4p_{x}^{2}\Delta-x\left(E^{2}-m^2\right)-\mu^{2}\frac{\left(x+a^{2}\right)\left(x+b^{2}\right)}{\Delta}\left(E+\frac{a}{x+a^{2}}\Phi+\frac{b}{x+b^{2}}\Psi\right)^{2}\nonumber\\
    &\qquad-\left(a^{2}-b^{2}\right)\left(\frac{\Phi^{2}}{x+a^{2}}-\frac{\Psi^{2}}{x+b^{2}}\right)\\
    \mathcal{H}_\theta&=p_{\theta}^{2}+\frac{\Phi^{2}}{\sin^{2}\theta}+\frac{\Psi^{2}}{\cos^{2}\theta}-\left(E^{2}-m^{2}\right)\left(a^{2}\cos^{2}\theta+b^{2}\sin^{2}\theta\right)\,.
\end{align}
Since $\mathcal{H}_x$ depends only on $x$, and $\mathcal{H}_\theta$ only on $\theta$, each of these functions must be constant to satisfy the Hamilton-Jacobi equation. In particular we write
\begin{equation}
    \mathcal{H}_x=-\mathcal{Q}+\frac{a^2+b^2}{2}\left(E^2-m^2\right)\, , \quad\mathcal{H}_\theta=\mathcal{Q}-\frac{a^2+b^2}{2}\left(E^2-m^2\right)\,.
\end{equation}
So the equations of motion read
\begin{align}
    \rho^{2}\dot{t}&=E\rho^{2}+\mu^{2}\frac{\left(x+a^{2}\right)\left(x+b^{2}\right)}{\Delta}\left(E+\frac{a}{x+a^{2}}\Phi+\frac{b}{x+b^{2}}\Psi\right)\,,\\\rho^{2}\dot{\phi}&=\frac{\Phi}{\sin^{2}\theta}-\mu^{2}a\frac{x+b^{2}}{\Delta}\left(E+\frac{a}{x+a^{2}}\Phi+\frac{b}{x+b^{2}}\Psi\right)-\frac{a^{2}-b^{2}}{x+a^{2}}\Phi\,,\\\rho^{2}\dot{\psi}&=\frac{\Psi}{\cos^{2}\theta}-\mu^{2}b\frac{x+a^{2}}{\Delta}\left(E+\frac{a}{x+a^{2}}\Phi+\frac{b}{x+b^{2}}\Psi\right)+\frac{a^{2}-b^{2}}{x+b^{2}}\Psi\,,\\\rho^{4}\dot{x}^{2}&=4\mathcal{X}\,,\\\rho^{4}\dot{\theta}^{2}&=\Theta\,,
\end{align}
with
\begin{align}
    \mathcal{X}&=\Delta\left[\left(E^{2}-m^{2}\right)\left(x+\frac{a^2 + b^2 }{2}\right)+\left(a^{2}-b^{2}\right)\left(\frac{\Phi^{2}}{x+a^{2}}-\frac{\Psi^{2}}{x+b^{2}}\right)-\mathcal{Q}\right]\nonumber\\
    &\qquad+\mu^{2}\left(x+a^{2}\right)\left(x+b^{2}\right)\left(E+\frac{a}{x+a^{2}}\Phi+\frac{b}{x+b^{2}}\Psi\right)^{2}\,,\\
    \Theta&=\mathcal{Q}-\frac{a^{2}-b^{2}}{2}\left(m^{2}-E^{2}\right)\cos\left(2\theta\right)-\frac{\Phi^{2}}{\sin^{2}\theta}-\frac{\Psi^{2}}{\cos^{2}\theta}\,.
\end{align}
Notice that the separation constant $\mathcal{Q}$ is not the usual one found
in the literature\cite{Frolov:2003en,Diemer:2014lba}, usually denoted
by $K$.

Previous work on radially bound timelike geodesics restricted the analysis to the $\theta=0$ and $\theta=\pi/2$ hypersurfaces, these are fixed points of the Killing vectors fields $\partial_\phi$ and $\partial_\psi$, respectively, thus the associated angular momentum must vanish. This work will not impose this restriction, thus both angular momenta can be non-zero giving rise to more complex dynamics. 

%%%%%%%%%%%%%%%%%%%%%%%%%%%%%%%%%%%%%%%%%%%%%%%%%%%%%%%%%%%%%%%
%%%%%%%%%%%%%%%%%%%%%%%%%%%%%%%%%%%%%%%%%%%%%%%%%%%%%%%%%%%%%%%
\section{No bound timelike geodesics\label{sec:The_Theorem_Time}}
%%%%%%%%%%%%%%%%%%%%%%%%%%%%%%%%%%%%%%%%%%%%%%%%%%%%%%%%%%%%%%%
%%%%%%%%%%%%%%%%%%%%%%%%%%%%%%%%%%%%%%%%%%%%%%%%%%%%%%%%%%%%%%%

This section will establish that timelike geodesics in the $d=5$ Myers Perry spacetime can never be \emph{radially} bounded outside the event horizon. No restrictions will be made to the latitudinal motion of the test particles in the following analysis. Hence, both the equations for latitudinal and radial motion will need to be considered. The derivation of this result is made easier with the introduction of the following parameters:
\begin{align}
p=a+b\,, & \quad q=a-b\,,\nonumber\\
\alpha=\Phi+\Psi\,, & \quad\beta=\Phi-\Psi\,.
\end{align}
In terms of these the horizon is located at
\begin{equation}
x_{H}=\frac{\mu^{2}-p^{2}+\mu^{2}-q^{2}}{4}+\frac{1}{2}\sqrt{\left(\mu^{2}-p^{2}\right)\left(\mu^{2}-q^{2}\right)}:=\frac{\mu_{p}^{2}+\mu_{q}^{2}}{4}+\frac{1}{2}\sqrt{\mu_{p}^{2}\mu_{q}^{2}}\,.
\end{equation}
Where $\mu_{p}^{2}=\mu^{2}-p^{2}$ and $\mu_{q}^{2}=\mu^{2}-q^{2}$.
Then it is clear that in order to have $x_{H}\geq0$ one must have $\mu_{p}^{2}\geq0$
and $\mu_{q}^{2}\geq0$.

%%%%%%%%%%%%%%%%%%%%%%%%%%%%%%%%%%%%%%%%%%%%%%%%%%%%%%%%%%%%%%%
%%%%%%%%%%%%%%%%%%%%%%%%%%%%%%%%%%%%%%%%%%%%%%%%%%%%%%%%%%%%%%%
\subsection{Latitudinal motion}
%%%%%%%%%%%%%%%%%%%%%%%%%%%%%%%%%%%%%%%%%%%%%%%%%%%%%%%%%%%%%%%
%%%%%%%%%%%%%%%%%%%%%%%%%%%%%%%%%%%%%%%%%%%%%%%%%%%%%%%%%%%%%%%

To study the motion along the latitude it is better to consider $u=\cos\left(2\theta\right)$, with $u\in\left[-1,1\right]$, as $\theta\in\left[0,\pi/2\right]$. Performing the change of variables, the equation of motion for $u$ reads:
\begin{equation}
    \rho^4\dot{u}^2=pq\frac{E^2-m^2}{2}u\left(1-u^2\right)+\left(1-u^2\right)Q-\left(\alpha^2+\beta^2+2\alpha \beta u\right):=\mathcal{U}\,.
\end{equation}
Notice that at the edges of the domain $\mathcal{U}$ is never positive, as:
\begin{equation}
    \mathcal{U}\left(-1\right)=-\left(\alpha-\beta\right)^2=-4\Psi^2\,,\quad \mathcal{U}\left(1\right)=-\left(\alpha\beta\right)^2=-4\Phi^2\,.
\end{equation}
In order to have physical motion along the latitude it is necessary that $\mathcal{U}\geq0$, for some range in $u\in\left[-1,1\right]$.

%%%%%%%%%%%%%%%%%%%%%%%%%%%%%%%%%%%%%%%%%%%%%%%%%%%%%%%%%%%%%%%
%%%%%%%%%%%%%%%%%%%%%%%%%%%%%%%%%%%%%%%%%%%%%%%%%%%%%%%%%%%%%%%
\subsection{Radial motion}
%%%%%%%%%%%%%%%%%%%%%%%%%%%%%%%%%%%%%%%%%%%%%%%%%%%%%%%%%%%%%%%
%%%%%%%%%%%%%%%%%%%%%%%%%%%%%%%%%%%%%%%%%%%%%%%%%%%%%%%%%%%%%%%

The radial motion is controlled by $4\mathcal{X}$ which can be written as a quadratic polynomial in the energy, $4\mathcal{X}=AE^{2}+BE+C$,
with coefficients
\begin{align}
A= & 4\left[\Delta\left(x+\mu^2+\frac{a^{2}+b^{2}}{2}\right)+\mu^{4}x\right]\,,\nonumber \\
B= & 8\mu^{2}\left[a\,\Phi\left(x+b^{2}\right)+b\,\Psi\left(x+a^{2}\right)\right]\,,\\
C= & 4\mu^{2}\left(b\Phi+a\Psi\right)^{2}+4\left(a^{2}-b^{2}\right)\left[\left(x+b^{2}\right)\Phi^{2}-\left(x+a^{2}\right)\Psi^{2}\right]-4\Delta\left[m^2\left(x+\frac{a^2+b^2}{2}\right)+Q\right]\,.\nonumber 
\end{align}
Notice that $A>0$, as $\Delta>0$ and $x>0$ outside the event horizon. The quadratic nature of this equation implies it can  be written in terms of effective potentials $V_{\pm}^{{\rm eff}}$~\cite{Wilkins:1972rs,Chandrasekhar:1985kt} as 
\begin{equation}
\rho^{4}\dot{x}^{2}=A\left(E-V_{+}^{{\rm eff}}\right)\left(E-V_{-}^{{\rm eff}}\right)\,,
\end{equation}
with 
\begin{equation}
V_{\pm}^{{\rm eff}}=\frac{-B\pm\sqrt{B^{2}-4AC}}{2A}\,.
\end{equation}
This way the radial turning points occur when $E=V_{\pm}^{{\rm eff}}\implies\mathcal{X}=0$.

Physical motion along the radial direction requires $\mathcal{X}\geq0$, therefore, since $A>0$,
the energy, $E$, must satisfy either $E\leq V_{-}^{{\rm eff}}$ or $E\geq V_{+}^{{\rm eff}}$. However, these conditions alone do not fully determine whether a trajectory is physically admissible. By analogy with four-dimensional spacetimes, one expects that physically meaningful trajectories should satisfy $E\geq 0$ at spatial infinity. It is important to emphasize that the conditions $E\leq V_{-}^{{\rm eff}}$ and $E\geq V_{+}^{{\rm eff}}$ are independent and do not necessarily imply $E\geq 0$ or $E\leq 0$, respectively. Depending on the choice of parameters, it is possible to have regions where $V_{+}^{\mathrm{eff}}<0$ or $V_{+}^{\mathrm{eff}}>0$. Meaning that these conditions are independent and must both be considered for a full analysis of radial motion. However performing the substitutions
\begin{equation}
\Phi\rightarrow-\Phi\,,\quad\Psi\rightarrow-\Psi\,,\quad E\rightarrow-E\,,
\end{equation}
one effectively transforms $V_{\pm}^{{\rm eff}}\rightarrow - V_{\mp}^{{\rm eff}}$,
thus one can consider only $E\geq V_{+}^{{\rm eff}}$ and then extend
the analysis to $E\leq V_{-}^{{\rm eff}}$ by means of the above transformation. 

To investigate the possibility of bending it is useful to study the
behaviour of the potentials at infinity:

\begin{equation}
\lim_{x\rightarrow\infty}V_{\pm}^{{\rm eff}}=\pm m\,.
\end{equation}
At infinity these effective potential should tend to the test particle's
mass, thus one gets the expected behaviour
for the potentials.

In order to have radially bounded motion one must have at least three
radial turning points, i.e. three distinct roots of the function $\mathcal{X}$,
outside the event horizon for any value of $E>0$, since $V_{+}^{{\rm eff}}\rightarrow m^{+}$
at infinity. This reasoning is illustrated in Fig. (\ref{fig:DummyPotentetial}),
where a dummy potential is depicted with three turning points for
some values of $E$.

\begin{center}
\begin{figure}
\begin{centering}
\includegraphics[width=0.7\textwidth]{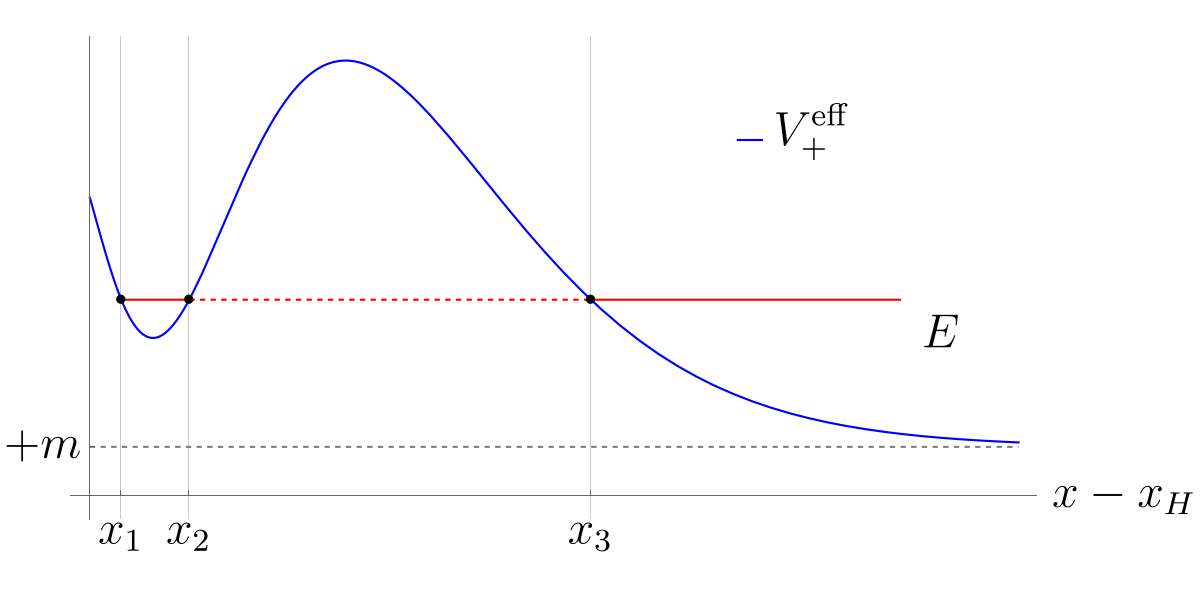}
\par\end{centering}
\caption{An illustrative example of an effective potential where bound orbits
are possible. This potential has three distinct roots, corresponding
to turning points, outside the event horizon, denoted $x_1\,,x_2$ and $x_3$. For a particle with energy $E$ bounded motion is possible
between $x_{1}$ and $x_{2}$, as depicted in the plot.\label{fig:DummyPotentetial}}
\end{figure}
\par\end{center}

Now, a new radial coordinate $y$ is introduced, defined by
\begin{equation}
x=\frac{y}{2}+\frac{\mu_{p}^{2}+\mu_{q}^{2}}{4}\,.
\end{equation}
Note that $y_{H}=\sqrt{\mu_{p}^{2}\mu_{q}^{2}}\geq0$ and $y>0$ outside
the event horizon. In terms of these new coordinates one obtains
\begin{equation}
4\mathcal{X}=\frac{E^{2}-m^2}{2}y^{3}+c_{2}y^{2}+c_{1}y+c_{0}\,.\label{eq:poly}
\end{equation}
With coefficients
\begin{align}
    c_{2}=&\mu^{2}E^{2}+\mu^{2}\frac{E^{2}-m^{2}}{2}-\mathcal{Q}\,,\nonumber\\
    c_{1}=&2\left(\mu^{2}E+\alpha p\right)\left(\mu^{2}E+\beta q\right)-\frac{E^{2}-m^{2}}{2}\mu_{q}^{2}\mu_{p}^{2}\,,\label{eq:coeffs} \\
    \nonumber c_{0}=&\mu_{q}^{2}\mu_{p}^{2}\left[\mathcal{Q}-2\mu^{2}E^{2}+\frac{\mu^{2}}{2}\left(E^{2}+m^{2}\right)\right]+\mu_{q}^{2}\left(\mu^{2}E+\alpha p\right)^{2}+\mu_{p}^{2}\left(\mu^{2}E+\beta q\right)^{2}\,.
\end{align}
Notice that the zeroth order term can be written in terms of $c_{2}$, as
\begin{equation}
    c_{0}=\mu_{q}^{2}\mu_{p}^{2}\left(-c_{2}\right)+\mu_{q}^{2}\left(\mu^{2}E+\alpha p\right)^{2}+\mu_{p}^{2}\left(\mu^{2}E+\beta q\right)^{2}\,.\label{eq:c0c2}
\end{equation}

Since, by construction $y>0$ outside the event horizon the existence
the bounded motion is related to the number of \emph{positive} roots
of Eq. (\ref{eq:poly}). To do so, it is useful to use the Descartes'
rule of signs, which relates the number of positive roots of a polynomial
with the number of sign changes of its coefficients.

\paragraph*{Descartes' rule of signs: }

if the non-zero terms of a single-variable polynomial with real coefficients
are ordered by descending variable exponent, then the number of positive
roots of the polynomial is either \emph{i)} equal to the number of
sign changes between consecutive (non-zero) coefficients, or \emph{ii)}
is less than it by an even number. In this context, a root of multiplicity
$k$ is counted as $k$ roots (see e.g. \cite{Anderson1998447}).\\

In order to have three positive roots of $\mathcal{X}$, the above rule states that three sign changes must occur, since $\mathcal{X}$ is a cubic polynomial this implies that all the coefficients must be non-zero, excluding the limiting case $E^2=m^2$, where the cubic coefficient vanishes. Thus there are two possibilities: 
\begin{align}
    (i) &\quad E^2>m^2\,,c_2<0\,,c_1>0\,,c_0<0\,,\\
    (ii) &\quad E^2<m^2\,,c_2>0\,,c_1<0\,,c_0>0\,.
\end{align}

Notice that possibility $(i)$ is impossible, since if $c_2<0$, then $c_0>0$, as it is a sum of all positive terms \ref{eq:c0c2}. Now possibility $(ii)$ will be considered, which implies that bounded motion requires $E^2<m^2$, as it happens in $4$ dimensional spacetimes.

The coefficient $c_1$ can be written as:
\begin{equation}
    c_1=2\left(\mu^{2}E+\alpha p\right)\left(\mu^{2}E+\beta q\right)-c_3\mu_{q}^{2}\mu_{p}^{2}\,,
\end{equation}
since $c_3<0$ and $\mu^2_{p,q}>0$ the last term on the right hand side is positive, then a necessary condition to have $c_1<0$ is $2\left(\mu^{2}E+\alpha p\right)\left(\mu^{2}E+\beta q\right)<0$. This expression is a convex parabola on $E$ with roots at $-\alpha p/\mu^2$ and $-\beta q /\mu^2$, so it negative between these two roots. In turn this implies that if there is a region where $c_1$ this region is contained in the interval between the aforementioned roots. 

Next consider the following combination $\mathcal{P}={\cal U}+\left(1-u^{2}\right)c_{2}$, which is independent of $\mathcal{Q}$. Since $c_2>0$ and physical motion along the latitude implies that $\mathcal{U}>0$ for some interval in $u$, the combination $\mathcal{P}$ must be non-negative somewhere within the available domain. The expression for $\mathcal{P}$ reads:
\begin{equation}
    {\cal P}=\frac{1}{2}\left(1-u^{2}\right)\left(3\mu^{2}+pqu\right)E^{2}-\left(\alpha^{2}+\beta^{2}+2u\alpha\beta\right)-\frac{m^{2}}{2}\left(1-u^{2}\right)\left(\mu^{2}+pqu\right)\,.
\end{equation}
This possesses no linear term on the energy, so it will have an extremum at $E=0$, moreover, the quadratic coefficient is positive as $1-u^2\in \left[0,1\right]$ and $3\mu^{2}+pqu\in\left[2\mu^2,4\mu^2\right]$. Therefore, $E=0$ is a minimum of $\mathcal{P}$, where
\begin{align}
    \mathcal{P}(0)&=-\left(\alpha^{2}+\beta^{2}+2u\alpha\beta\right)-\frac{m^{2}}{2}\left(1-u^{2}\right)\left(\mu^{2}+pqu\right)\nonumber\\
    &=-\frac{\left(1+u\right)\left(\alpha+\beta\right)^2+\left(1-u\right)\left(\alpha-\beta\right)^2}{2}-\frac{m^{2}}{2}\left(1-u^{2}\right)\left(\mu^{2}+pqu\right)\,.
\end{align}
For $u\in\left[-1,1\right]$ this is always negative. Since $\mathcal{P}$ is symmetric with respect to $E=0$, where it is negative, a necessary condition for it to be positive for some intervals in $E\in\left[-m,m\right]$, is that $\mathcal{P}(-m)=\mathcal{P}(m)>0$, that is
\begin{equation}
    \mathcal{P}(m)=-\left(\alpha^{2}+\beta^{2}+2u\alpha\beta\right)+m^{2}\mu^{2}\left(1-u^{2}\right)>0\,.
\end{equation}
This is, in turn a polynomial in $u$ which is non-positive at $u=\pm1$, then since it is concave (the quadratic coefficient is $-m^2\mu^2<0$) for it to be positive in some range its maximum value must be obtained for $u=u_{m}=-\alpha\beta/\mu^2m^2\in\left[-1,1\right]$ and ${\cal P}\left(E=m\,,u=u_{m}\right)>0$, these yield the following system of inequalities:
\begin{equation}
\begin{cases}
\frac{\left(\alpha^{2}-\mu^{2}m^{2}\right)\left(\beta^{2}-\mu^{2}m^{2}\right)}{\mu^{2}m^{2}}>0\\
-1<-\frac{\alpha\beta}{\mu^{2}m^{2}}<1
\end{cases}
\Leftrightarrow
\begin{cases}
\alpha^{2}<\mu^{2}m^{2}\\
\beta^{2}<\mu^{2}m^{2}
\end{cases}\,.
\end{equation}

Now to assert the impossibility of having radially bound timelike geodesics in the $d=5$ MP spacetime it will be proven that it is impossible to simultaneously satisfy $c_1<0$ and $\mathcal{P}>0$. Since $\mathcal{P}$ is a convex parabola and $c_1$ can only be negative for energy $E$ between $-\alpha p/\mu^2$ and $-\beta q /\mu^2$, then $\mathcal{P}$ must be positive at at least one of these points, otherwise it is negative in the entire interval violating the necessary condition for radially bound motion. At $E=-\alpha p/\mu^{2}$ one has
\begin{equation}
    {\cal P}\left(-\alpha p/\mu^{2}\right)=\underbrace{\left(1-u^{2}\right)\left(\mu^{2}+pqu\right)\frac{E^2-m^2}{2}}_{<0}+\underbrace{\frac{\alpha^{2}p^{2}}{\mu^{2}}\left(1-u^{2}\right)-\left(\alpha^{2}+\beta^{2}+2u\alpha\beta\right)}_{:=\mathcal{A}}\,.
\end{equation}
Since $E^2<m^2$ and $-\mu<p,q<\mu$ the first term on the right hand side of the above equation is negative, as indicated. So a necessary (but not sufficient) condition to have ${\cal P}\left(-\alpha p/\mu^{2}\right)>0$ is that the remaining terms, denoted $\mathcal{A}$, are positive
\begin{equation}
    {\cal A}=-\alpha^{2}\left[1-\frac{p^{2}}{\mu^{2}}\left(1-u^{2}\right)\right]-2u\alpha\beta-\beta^{2}
\end{equation}
this is a concave parabola in $\alpha$, as the quadratic coefficient is negative since $0<p^2/\mu^2<1$, so for this expression to be positive, it has to have a maximum $\alpha_{m}\in\left[-\mu\, m,\mu\, m\right]$, and the maximum must be positive, we obtain
\begin{align}
    \alpha_{m}&=-\mu^{2}\frac{u\beta}{\mu^{2}-\left(1-u^{2}\right)p^{2}}\,,\\
    \mathcal{A}\left(\alpha_{m}\right)&=-\left(1-u^{2}\right)\frac{\mu_p^2}{\mu_p^{2}+u^{2}p^{2}}\beta^{2}\,.
\end{align}
From this it is clear that $\mathcal{A}\left(\alpha_{m}\right)<0$. A similar reasoning for $E=-\beta q/\mu^2$ yields $\mathcal{P}\left(-\beta q/\mu^2\right)<0$. With this it is possible to conclude that if $c_1<0$, then one necessarily has $\mathcal{P}<0$.

The discussion above is illustrated in Fig. (\ref{fig:Contradiction}), where $c_1$ and $\mathcal{P}$ are plotted for some values of the parameters. There it is possible to see that it is, in fact, impossible to have $c_1<0$ and $\mathcal{P}>0$. This contradiction between the necessary conditions for radially bound timelike geodesics implies that such geodesics cannot occur in the $d=5$ MP spacetime.

\begin{center}
\begin{figure}
\begin{centering}
\includegraphics[width=0.6\textwidth]{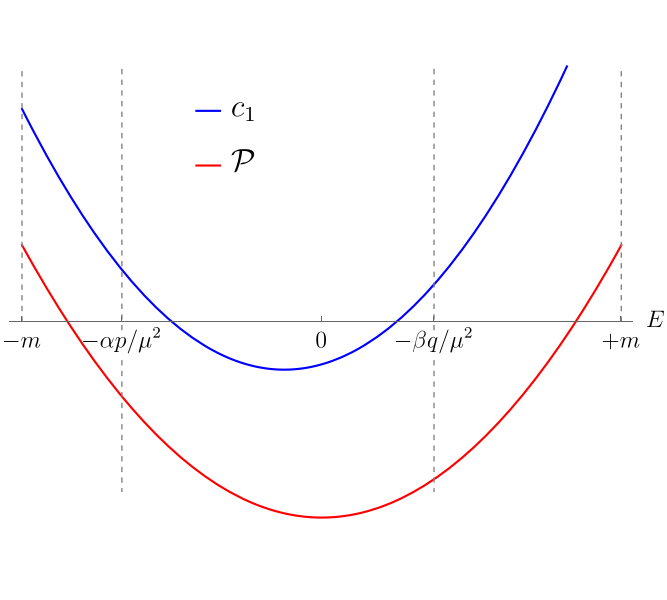}
\par\end{centering}
\caption{The polynomials $\mathcal{P}$ and $c_1$, for some values of the parameters. Here it is possible to appreciate that $\mathcal{P}$ is negative at the origin, and thus for it to be positive in some range contained in $E\in[-m,m]$ it has to be positive at $E=\pm m $. Moreover the coefficient $c_1$ is always plotted, as discussed in the text this can only be negative in some range between $-\alpha p/\mu^2$ and $-\beta q/\mu^2$. Since, as it is shown, $\mathcal{P}$ is always negative at these value of the energy it is impossible to have $c_1<0$ and $\mathcal{P}>0$, reaching a contradiction between the conditions for radially bound timelike geodesics. Thus, this establishes that such geodesics are not possible in the $d=5$ MP spacetime. \label{fig:Contradiction}}
\end{figure}
\par\end{center}

%%%%%%%%%%%%%%%%%%%%%%%%%%%%%%%%%%%%%%%%%%%%%%%%%%%%%%%%%%%%%%%
%%%%%%%%%%%%%%%%%%%%%%%%%%%%%%%%%%%%%%%%%%%%%%%%%%%%%%%%%%%%%%%
\section{No bound null geodesics\label{sec:The_Theorem}}
%%%%%%%%%%%%%%%%%%%%%%%%%%%%%%%%%%%%%%%%%%%%%%%%%%%%%%%%%%%%%%%
%%%%%%%%%%%%%%%%%%%%%%%%%%%%%%%%%%%%%%%%%%%%%%%%%%%%%%%%%%%%%%%

In this section we will consider the particular case of $m=0$, \emph{i.e.} if it is possible for null geodesics to be radially bound. Setting $m=0$ in the radial polynomial still yields a quadratic polynomial in the energy, with $A>0$. Therefore, the reasoning in the previous section remains valid, the only major difference is the asymptotic behaviour of the effective potential, which is now:
\begin{equation}
\lim_{x\rightarrow\infty}V_{\pm}^{{\rm eff}}=\lim_{x\rightarrow\infty}\pm\sqrt{\frac{\mathcal{Q}}{x}}=0^{\pm}\,.
\end{equation}

Like in the timelike case radially bounded motion requires at least three radial turning points outside the event horizon for any value of $E>0$, since $V_{+}^{{\rm eff}}\rightarrow0^{+}$ at infinity.

The polynomial $4\mathcal{X}$, in terms of $y$, now reads:
\begin{equation}
4\mathcal{X}=\frac{E^{2}}{2}y^{3}+c_{2}y^{2}+c_{1}y+c_{0}\,.\label{eq:poly}
\end{equation}
Where the remaining coefficients are obtained by setting $m=0$ in (\ref{eq:coeffs}), the main difference is that now the cubic coefficient can never be negative, hence the only way to have three distinct roots is if:
\begin{equation}
    E^2>0\,,c_{2}<0\,,c_{1}>0\,,c_{0}<0\,.
\end{equation}
However, Eq. (\ref{eq:c0c2}) which relates $c_2$ and $c_0$ still holds, meaning that it is impossible to have simultaneously $c_2<0$ and $c_0<0$. Therefore, one concludes that there are no bound null orbits in the Myers-Perry spacetime. An example of the effective potential $V_{+}^{{\rm eff}}$ for a given choice of parameters is presented in Fig. (\ref{fig:Example}), for that choice of parameters it becomes clear that there can be zero, one or two roots, depending on the photon's energy. The possibility of a root at $y=0$ is discussed below.
\begin{center}
\begin{figure}
\begin{centering}
\includegraphics[width=0.7\textwidth]{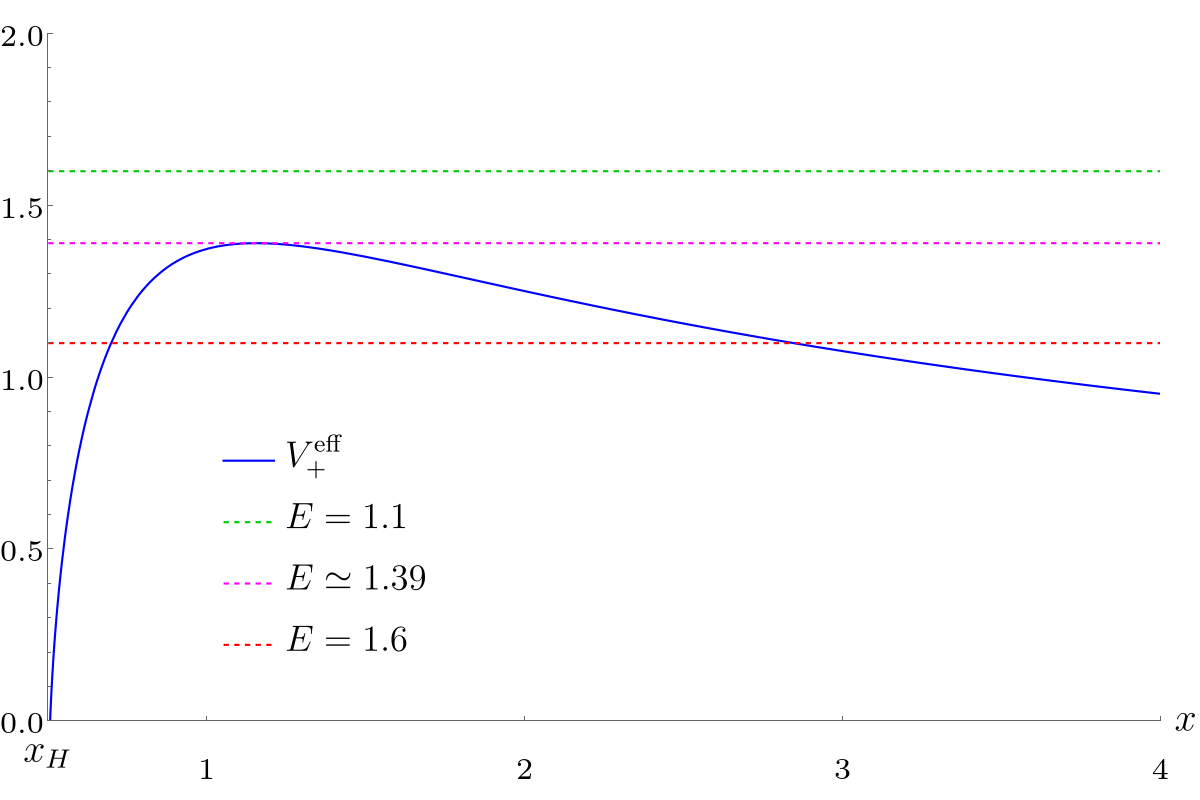}
\par\end{centering}
\caption{The effective potential for $Q=2\,,\alpha=4\,,\beta=-6\,,\mu_{p}^{2}=1/4\,,\mu_{q}^{2}=9/4-\sqrt{2}$.
This choice of parameters satisfies the light ring condition $Q=\alpha^{2}$,
so the maximum of the potential corresponds a light ring. Three different
possible values for the energy are depicted, one without turning points,
another with only one (corresponding to the light ring) and another
one with two. \label{fig:Example}}
\end{figure}
\par\end{center}

\section{A caveat - the extremal limit}

In general, the horizon is located at some $y_{H}>0$, so there is
no need to worry about a possible root at $y=0$ for the polynomials governing the radial motion of both null and timelike geodesics~\footnote{Note that a similar behaviour also occurs in the extremal Kerr spacetime where the radial location of the inner light ring coincides with that of the horizon, although they are at some finite proper distance.}, which is not covered by the previous analyses. This possibility is discussed in this section. In order to have $y_{H}=0$ one needs $\mu_{q}^{2}=0$ or $\mu_{p}^{2}=0$, which corresponds to an extremal
black hole as the surface gravity of the MP black hole,
\begin{equation}
\kappa=\frac{1}{\mu^{2}}\sqrt{\frac{\mu_{p}^{2}\mu_{q}^{2}}{\mu_{p}^{2}+\mu_{q}^{2}+2\sqrt{\mu_{p}^{2}\mu_{q}^{2}}}}\,,
\end{equation}
vanished in this case.

A necessary and sufficient condition for a root of $\mathcal{X}$ at $y=0$ is that
$c_{0}=0$, this condition is only met if and only if $\mu_{q}^{2}=0$ and $\mu_{p}^{2}=0$
simultaneously, where there is a factorisation and the expression simplifies to
\begin{equation}
4\mathcal{X}=y\left(\frac{E^{2}-m^2}{2}y^{2}+c_2y+c_1\right)\,,
\end{equation}
where $m$ can be straightly set to zero to obtain the null case.

This particular limit, corresponding to an extremal single rotating
black hole, where one of the spin parameters vanishes and the other
extremises the solution, \emph{i.e} either $a=0\,,|b|=\mu$ or $b=0\,,|a|=\mu$. However, for such values of the parameters
the metric (\ref{eq:MP_line}) possesses a naked singularity at the
horizon, this can be appreciated by considering the general expression
for the Kretschmann scalar

\begin{equation}
R_{\mu\nu\rho\sigma}R^{\mu\nu\rho\sigma}=\frac{24\mu^{4}}{\rho^{12}}\left(4x-3\rho^{2}\right)\left(4x-\rho^{2}\right)\,.\label{eq:Kretshcmann}
\end{equation}
Considering the case $a^{2}=\mu^{2}$ and $b=0$ the Kretschmann scalar
at the horizon is 

\begin{equation}
\left.R_{\mu\nu\rho\sigma}R^{\mu\nu\rho\sigma}\right|_{x=0}=\frac{72}{\mu^{4}\cos^{8}\theta}\,,
\end{equation}
this is clearly singular when $\theta=\pi/2$, which is inside the
range for that coordinate. If $a=0$ and $b^{2}=\mu^{2}$ the cosine
becomes a sine and the expression is singular when $\theta=0$. Therefore,
as only solutions without naked singularities are of interest this case will lie outside the range of the theorem established in this letter, that states 

\section{Conclusions and discussion\label{sec:Conclusions}}

The equations of motion for particles in the background of the general $d=5$ MP metric have been known for several years~\cite{Frolov:2003en}. These equations have been extensively used in attempts to generalize the result that no timelike bound geodesics can exist outside the event horizon of the $d=5$ Tangherlini black hole~\cite{Tangherlini:1963bw} to its spinning counterpart. The $d=5$ MP black hole rotates on two independent planes, yielding two independent angular momenta, this gives rise to a very rich and complex geodesic structure. To simplify the analysis, previous works restricted their studies to what are typically referred to as the ``equatorial planes'' of the spacetime~\cite{Cardoso:2008bp,Dadhich:2020svz,Dadhich:2021vdd}. However, these do not correspond to fixed points of a $\mathbb{Z}_2$ symmetry, as in the 4D Kerr case, but are instead fixed points of the rotational symmetries~\cite{Myers:1986un,Myers:2011yc}, making them more akin to a rotation axis.  The study presented in this paper goes beyond this simplification and considers geodesics that can exhibit motion along $\theta$. While this broader approach reproduces the previous results stating that radially bound timelike geodesics are not possible, it represents, as far as the author is aware, the first complete proof that extends beyond the usual $\theta = 0$ or $\pi/2$ hyperplanes.

Null geodesics were also analysed. Although their analytical solutions are available in the literature~\cite{Diemer:2014lba}, the possibility of radially bound null geodesics had not been yet discussed. This was achieved in this paper by following a strategy similar to the one used by Wilkins~\cite{Wilkins:1972rs} to prove that no bound null geodesics exist in the Kerr spacetime. The result was a theorem establishing that it is impossible to have such geodesics in the $d=5$ MP spacetime. This has possible implications on the stability of the $d=5$ MP black hole, as there are no orbits where energy is expected to accumulate \cite{Cunha:2024ajc}. Moreover, it implies that the hypershadow of the general $d=5$ MP black hole is determined by the unstable spherical photon orbits. 

The main result of this paper can be summarized in the following theorem:
\paragraph*{Theorem} the spacetime outside the event horizon of a general $d=5$ MP solution, regular on and outside the event horizon, does not allow for radially bound null geodesics.\\

A corollary of this theorem is that there are no stable spherical (constant radial coordinate) orbits, whether null or timelike.

These findings provide a comprehensive understanding of the geodesic structure in the $d=5$ MP black hole spacetime, emphasizing the robustness of higher-dimensional rotating black holes against instabilities. In 4 spacetime dimensions it is known that vacuum black holes always possess a topologically spherical event horizon~\cite{Hawking:1971vc,Hawking:1973uf}, this is no longer the case in 5 dimensions, where it is possible to have black hole solutions with topologically toroidal event horizons~\cite{Emparan:2001wn}. Therefore an interesting follow up on this work would be to see if such black ring spacetimes can harbour radially bound timelike geodesics. 

Another interesting extension of our analysis is the case of charged rotating black holes. It is easy to verify that the addition of charge does alter the possibility of radially bound geodesics in static five-dimensional black hole spacetimes. However, whether charge could qualitatively change the behaviour in the spinning case remains an open question. Unfortunately, no fully analytical solution for a charged rotating Myers-Perry black hole in five dimensions is known. While a perturbative slowly spinning solution exists \cite{Aliev:2006yk}, it does not admit full separability of the geodesic equations and lacks the Carter-like constant, implying that the motion is not integrable. As a result, our analytical approach cannot be directly applied to this case. Addressing this question would require numerical methods to analyse geodesic motion in such spacetimes, which lies beyond the scope of this work but is an interesting avenue for future investigation.

\begin{acknowledgments}
The author grateful to C. Herdeiro and P. Cunha for their valuable comments and suggestions during the writing of the manuscript. This work is supported by the Center for Research and Development in Mathematics and Applications (CIDMA) through the Portuguese Foundation for Science and Technology (FCT -- Fundaç\~ao para a Ci\^encia e a Tecnologia) through projects: UIDB/04106/2020, PTDC/FIS-AST/3041/2020, 2022.04560.PTDC (\url{https://doi.org/10.54499/UIDB/04106/2020}; \url{https://doi.org/10.54499/UIDP/04106/2020}; and \url{http://doi.org/10.54499/PTDC/FIS-AST/3041/2020}; \ and \url{https://doi.org/10.54499/2022.04560.PTDC}), additionally this work was supported by the project 2024.05617.CERN. This work has further been supported by the European Horizon Europe staff exchange (SE) programme HORIZON-MSCA-2021-SE-01 Grant No.\ NewFunFiCO-101086251. J.N. is supported by the FCT grant 10.54499/2021.06539.BD (\url{https://doi.org/10.54499/2021.06539.BD}). 
\end{acknowledgments}

\bibliography{apssamp}% Produces the bibliography via BibTeX.

\begin{thebibliography}{29}%
\makeatletter
\providecommand \@ifxundefined [1]{%
 \@ifx{#1\undefined}
}%
\providecommand \@ifnum [1]{%
 \ifnum #1\expandafter \@firstoftwo
 \else \expandafter \@secondoftwo
 \fi
}%
\providecommand \@ifx [1]{%
 \ifx #1\expandafter \@firstoftwo
 \else \expandafter \@secondoftwo
 \fi
}%
\providecommand \natexlab [1]{#1}%
\providecommand \enquote  [1]{``#1''}%
\providecommand \bibnamefont  [1]{#1}%
\providecommand \bibfnamefont [1]{#1}%
\providecommand \citenamefont [1]{#1}%
\providecommand \href@noop [0]{\@secondoftwo}%
\providecommand \href [0]{\begingroup \@sanitize@url \@href}%
\providecommand \@href[1]{\@@startlink{#1}\@@href}%
\providecommand \@@href[1]{\endgroup#1\@@endlink}%
\providecommand \@sanitize@url [0]{\catcode `\\12\catcode `\$12\catcode `\&12\catcode `\#12\catcode `\^12\catcode `\_12\catcode `\%12\relax}%
\providecommand \@@startlink[1]{}%
\providecommand \@@endlink[0]{}%
\providecommand \url  [0]{\begingroup\@sanitize@url \@url }%
\providecommand \@url [1]{\endgroup\@href {#1}{\urlprefix }}%
\providecommand \urlprefix  [0]{URL }%
\providecommand \Eprint [0]{\href }%
\providecommand \doibase [0]{https://doi.org/}%
\providecommand \selectlanguage [0]{\@gobble}%
\providecommand \bibinfo  [0]{\@secondoftwo}%
\providecommand \bibfield  [0]{\@secondoftwo}%
\providecommand \translation [1]{[#1]}%
\providecommand \BibitemOpen [0]{}%
\providecommand \bibitemStop [0]{}%
\providecommand \bibitemNoStop [0]{.\EOS\space}%
\providecommand \EOS [0]{\spacefactor3000\relax}%
\providecommand \BibitemShut  [1]{\csname bibitem#1\endcsname}%
\let\auto@bib@innerbib\@empty
%</preamble>
\bibitem [{\citenamefont {Wilkins}(1972)}]{Wilkins:1972rs}%
  \BibitemOpen
  \bibfield  {author} {\bibinfo {author} {\bibfnamefont {D.~C.}\ \bibnamefont {Wilkins}},\ }\bibfield  {title} {\bibinfo {title} {{Bound Geodesics in the Kerr Metric}},\ }\href {https://doi.org/10.1103/PhysRevD.5.814} {\bibfield  {journal} {\bibinfo  {journal} {Phys. Rev. D}\ }\textbf {\bibinfo {volume} {5}},\ \bibinfo {pages} {814} (\bibinfo {year} {1972})}\BibitemShut {NoStop}%
\bibitem [{\citenamefont {Tangherlini}(1963)}]{Tangherlini:1963bw}%
  \BibitemOpen
  \bibfield  {author} {\bibinfo {author} {\bibfnamefont {F.~R.}\ \bibnamefont {Tangherlini}},\ }\bibfield  {title} {\bibinfo {title} {{Schwarzschild field in n dimensions and the dimensionality of space problem}},\ }\href {https://doi.org/10.1007/BF02784569} {\bibfield  {journal} {\bibinfo  {journal} {Nuovo Cim.}\ }\textbf {\bibinfo {volume} {27}},\ \bibinfo {pages} {636} (\bibinfo {year} {1963})}\BibitemShut {NoStop}%
\bibitem [{\citenamefont {Emparan}\ and\ \citenamefont {Reall}(2002)}]{Emparan:2001wn}%
  \BibitemOpen
  \bibfield  {author} {\bibinfo {author} {\bibfnamefont {R.}~\bibnamefont {Emparan}}\ and\ \bibinfo {author} {\bibfnamefont {H.~S.}\ \bibnamefont {Reall}},\ }\bibfield  {title} {\bibinfo {title} {{A Rotating black ring solution in five-dimensions}},\ }\href {https://doi.org/10.1103/PhysRevLett.88.101101} {\bibfield  {journal} {\bibinfo  {journal} {Phys. Rev. Lett.}\ }\textbf {\bibinfo {volume} {88}},\ \bibinfo {pages} {101101} (\bibinfo {year} {2002})},\ \Eprint {https://arxiv.org/abs/hep-th/0110260} {arXiv:hep-th/0110260} \BibitemShut {NoStop}%
\bibitem [{\citenamefont {Emparan}\ and\ \citenamefont {Reall}(2008)}]{Emparan:2008eg}%
  \BibitemOpen
  \bibfield  {author} {\bibinfo {author} {\bibfnamefont {R.}~\bibnamefont {Emparan}}\ and\ \bibinfo {author} {\bibfnamefont {H.~S.}\ \bibnamefont {Reall}},\ }\bibfield  {title} {\bibinfo {title} {{Black Holes in Higher Dimensions}},\ }\href {https://doi.org/10.12942/lrr-2008-6} {\bibfield  {journal} {\bibinfo  {journal} {Living Rev. Rel.}\ }\textbf {\bibinfo {volume} {11}},\ \bibinfo {pages} {6} (\bibinfo {year} {2008})},\ \Eprint {https://arxiv.org/abs/0801.3471} {arXiv:0801.3471 [hep-th]} \BibitemShut {NoStop}%
\bibitem [{\citenamefont {Frolov}\ and\ \citenamefont {Stojkovic}(2003)}]{Frolov:2003en}%
  \BibitemOpen
  \bibfield  {author} {\bibinfo {author} {\bibfnamefont {V.~P.}\ \bibnamefont {Frolov}}\ and\ \bibinfo {author} {\bibfnamefont {D.}~\bibnamefont {Stojkovic}},\ }\bibfield  {title} {\bibinfo {title} {{Particle and light motion in a space-time of a five-dimensional rotating black hole}},\ }\href {https://doi.org/10.1103/PhysRevD.68.064011} {\bibfield  {journal} {\bibinfo  {journal} {Phys. Rev. D}\ }\textbf {\bibinfo {volume} {68}},\ \bibinfo {pages} {064011} (\bibinfo {year} {2003})},\ \Eprint {https://arxiv.org/abs/gr-qc/0301016} {arXiv:gr-qc/0301016} \BibitemShut {NoStop}%
\bibitem [{\citenamefont {Cardoso}\ \emph {et~al.}(2009)\citenamefont {Cardoso}, \citenamefont {Miranda}, \citenamefont {Berti}, \citenamefont {Witek},\ and\ \citenamefont {Zanchin}}]{Cardoso:2008bp}%
  \BibitemOpen
  \bibfield  {author} {\bibinfo {author} {\bibfnamefont {V.}~\bibnamefont {Cardoso}}, \bibinfo {author} {\bibfnamefont {A.~S.}\ \bibnamefont {Miranda}}, \bibinfo {author} {\bibfnamefont {E.}~\bibnamefont {Berti}}, \bibinfo {author} {\bibfnamefont {H.}~\bibnamefont {Witek}},\ and\ \bibinfo {author} {\bibfnamefont {V.~T.}\ \bibnamefont {Zanchin}},\ }\bibfield  {title} {\bibinfo {title} {{Geodesic stability, Lyapunov exponents and quasinormal modes}},\ }\href {https://doi.org/10.1103/PhysRevD.79.064016} {\bibfield  {journal} {\bibinfo  {journal} {Phys. Rev. D}\ }\textbf {\bibinfo {volume} {79}},\ \bibinfo {pages} {064016} (\bibinfo {year} {2009})},\ \Eprint {https://arxiv.org/abs/0812.1806} {arXiv:0812.1806 [hep-th]} \BibitemShut {NoStop}%
\bibitem [{\citenamefont {Dadhich}\ and\ \citenamefont {Shaymatov}(2022{\natexlab{a}})}]{Dadhich:2020svz}%
  \BibitemOpen
  \bibfield  {author} {\bibinfo {author} {\bibfnamefont {N.}~\bibnamefont {Dadhich}}\ and\ \bibinfo {author} {\bibfnamefont {S.}~\bibnamefont {Shaymatov}},\ }\bibfield  {title} {\bibinfo {title} {{On black hole formation in higher dimensions}},\ }\href {https://doi.org/10.1142/S0218271821501200} {\bibfield  {journal} {\bibinfo  {journal} {Int. J. Mod. Phys. D}\ }\textbf {\bibinfo {volume} {31}},\ \bibinfo {pages} {2150120} (\bibinfo {year} {2022}{\natexlab{a}})},\ \Eprint {https://arxiv.org/abs/2009.10528} {arXiv:2009.10528 [gr-qc]} \BibitemShut {NoStop}%
\bibitem [{\citenamefont {Dadhich}\ and\ \citenamefont {Shaymatov}(2022{\natexlab{b}})}]{Dadhich:2021vdd}%
  \BibitemOpen
  \bibfield  {author} {\bibinfo {author} {\bibfnamefont {N.}~\bibnamefont {Dadhich}}\ and\ \bibinfo {author} {\bibfnamefont {S.}~\bibnamefont {Shaymatov}},\ }\bibfield  {title} {\bibinfo {title} {{Circular orbits around higher dimensional Einstein and pure Gauss\textendash{}Bonnet rotating black holes}},\ }\href {https://doi.org/10.1016/j.dark.2022.100986} {\bibfield  {journal} {\bibinfo  {journal} {Phys. Dark Univ.}\ }\textbf {\bibinfo {volume} {35}},\ \bibinfo {pages} {100986} (\bibinfo {year} {2022}{\natexlab{b}})},\ \Eprint {https://arxiv.org/abs/2104.00427} {arXiv:2104.00427 [gr-qc]} \BibitemShut {NoStop}%
\bibitem [{\citenamefont {Bardeen}(1973)}]{Bardeen:1973tla}%
  \BibitemOpen
  \bibfield  {author} {\bibinfo {author} {\bibfnamefont {J.~M.}\ \bibnamefont {Bardeen}},\ }\bibfield  {title} {\bibinfo {title} {{Timelike and null geodesics in the Kerr metric}},\ }\href@noop {} {\bibfield  {journal} {\bibinfo  {journal} {Proceedings, Ecole d'Et\'e de Physique Th\'eorique: Les Astres Occlus : Les Houches, France, August, 1972, 215-240}\ ,\ \bibinfo {pages} {215}} (\bibinfo {year} {1973})}\BibitemShut {NoStop}%
\bibitem [{\citenamefont {Falcke}\ \emph {et~al.}(2000)\citenamefont {Falcke}, \citenamefont {Melia},\ and\ \citenamefont {Agol}}]{Falcke:1999pj}%
  \BibitemOpen
  \bibfield  {author} {\bibinfo {author} {\bibfnamefont {H.}~\bibnamefont {Falcke}}, \bibinfo {author} {\bibfnamefont {F.}~\bibnamefont {Melia}},\ and\ \bibinfo {author} {\bibfnamefont {E.}~\bibnamefont {Agol}},\ }\bibfield  {title} {\bibinfo {title} {{Viewing the shadow of the black hole at the galactic center}},\ }\href {https://doi.org/10.1086/312423} {\bibfield  {journal} {\bibinfo  {journal} {Astrophys. J. Lett.}\ }\textbf {\bibinfo {volume} {528}},\ \bibinfo {pages} {L13} (\bibinfo {year} {2000})},\ \Eprint {https://arxiv.org/abs/astro-ph/9912263} {arXiv:astro-ph/9912263} \BibitemShut {NoStop}%
\bibitem [{\citenamefont {{Goebel}}(1972)}]{Goebel}%
  \BibitemOpen
  \bibfield  {author} {\bibinfo {author} {\bibfnamefont {C.~J.}\ \bibnamefont {{Goebel}}},\ }\bibfield  {title} {\bibinfo {title} {{Comments on the ``vibrations'' of a Black Hole.}},\ }\href {https://doi.org/10.1086/180898} {\bibfield  {journal} {\bibinfo  {journal} {"The Astrophysical Journal"}\ }\textbf {\bibinfo {volume} {172}},\ \bibinfo {pages} {L95} (\bibinfo {year} {1972})}\BibitemShut {NoStop}%
\bibitem [{\citenamefont {Cardoso}\ \emph {et~al.}(2016)\citenamefont {Cardoso}, \citenamefont {Franzin},\ and\ \citenamefont {Pani}}]{Cardoso:2016rao}%
  \BibitemOpen
  \bibfield  {author} {\bibinfo {author} {\bibfnamefont {V.}~\bibnamefont {Cardoso}}, \bibinfo {author} {\bibfnamefont {E.}~\bibnamefont {Franzin}},\ and\ \bibinfo {author} {\bibfnamefont {P.}~\bibnamefont {Pani}},\ }\bibfield  {title} {\bibinfo {title} {{Is the gravitational-wave ringdown a probe of the event horizon?}},\ }\href {https://doi.org/10.1103/PhysRevLett.116.171101} {\bibfield  {journal} {\bibinfo  {journal} {Phys. Rev. Lett.}\ }\textbf {\bibinfo {volume} {116}},\ \bibinfo {pages} {171101} (\bibinfo {year} {2016})},\ \bibinfo {note} {[Erratum: Phys.Rev.Lett. 117, 089902 (2016)]},\ \Eprint {https://arxiv.org/abs/1602.07309} {arXiv:1602.07309 [gr-qc]} \BibitemShut {NoStop}%
\bibitem [{\citenamefont {Cardoso}\ \emph {et~al.}(2014)\citenamefont {Cardoso}, \citenamefont {Crispino}, \citenamefont {Macedo}, \citenamefont {Okawa},\ and\ \citenamefont {Pani}}]{Cardoso:2014sna}%
  \BibitemOpen
  \bibfield  {author} {\bibinfo {author} {\bibfnamefont {V.}~\bibnamefont {Cardoso}}, \bibinfo {author} {\bibfnamefont {L.~C.~B.}\ \bibnamefont {Crispino}}, \bibinfo {author} {\bibfnamefont {C.~F.~B.}\ \bibnamefont {Macedo}}, \bibinfo {author} {\bibfnamefont {H.}~\bibnamefont {Okawa}},\ and\ \bibinfo {author} {\bibfnamefont {P.}~\bibnamefont {Pani}},\ }\bibfield  {title} {\bibinfo {title} {{Light rings as observational evidence for event horizons: long-lived modes, ergoregions and nonlinear instabilities of ultracompact objects}},\ }\href {https://doi.org/10.1103/PhysRevD.90.044069} {\bibfield  {journal} {\bibinfo  {journal} {Phys. Rev. D}\ }\textbf {\bibinfo {volume} {90}},\ \bibinfo {pages} {044069} (\bibinfo {year} {2014})},\ \Eprint {https://arxiv.org/abs/1406.5510} {arXiv:1406.5510 [gr-qc]} \BibitemShut {NoStop}%
\bibitem [{\citenamefont {Keir}(2016)}]{Keir:2014oka}%
  \BibitemOpen
  \bibfield  {author} {\bibinfo {author} {\bibfnamefont {J.}~\bibnamefont {Keir}},\ }\bibfield  {title} {\bibinfo {title} {{Slowly decaying waves on spherically symmetric spacetimes and ultracompact neutron stars}},\ }\href {https://doi.org/10.1088/0264-9381/33/13/135009} {\bibfield  {journal} {\bibinfo  {journal} {Class. Quant. Grav.}\ }\textbf {\bibinfo {volume} {33}},\ \bibinfo {pages} {135009} (\bibinfo {year} {2016})},\ \Eprint {https://arxiv.org/abs/1404.7036} {arXiv:1404.7036 [gr-qc]} \BibitemShut {NoStop}%
\bibitem [{\citenamefont {Cunha}\ \emph {et~al.}(2023)\citenamefont {Cunha}, \citenamefont {Herdeiro}, \citenamefont {Radu},\ and\ \citenamefont {Sanchis-Gual}}]{Cunha:2022gde}%
  \BibitemOpen
  \bibfield  {author} {\bibinfo {author} {\bibfnamefont {P.~V.~P.}\ \bibnamefont {Cunha}}, \bibinfo {author} {\bibfnamefont {C.}~\bibnamefont {Herdeiro}}, \bibinfo {author} {\bibfnamefont {E.}~\bibnamefont {Radu}},\ and\ \bibinfo {author} {\bibfnamefont {N.}~\bibnamefont {Sanchis-Gual}},\ }\bibfield  {title} {\bibinfo {title} {{Exotic Compact Objects and the Fate of the Light-Ring Instability}},\ }\href {https://doi.org/10.1103/PhysRevLett.130.061401} {\bibfield  {journal} {\bibinfo  {journal} {Phys. Rev. Lett.}\ }\textbf {\bibinfo {volume} {130}},\ \bibinfo {pages} {061401} (\bibinfo {year} {2023})},\ \Eprint {https://arxiv.org/abs/2207.13713} {arXiv:2207.13713 [gr-qc]} \BibitemShut {NoStop}%
\bibitem [{\citenamefont {Novo}\ \emph {et~al.}(2024)\citenamefont {Novo}, \citenamefont {Cunha},\ and\ \citenamefont {Herdeiro}}]{Novo:2024wyn}%
  \BibitemOpen
  \bibfield  {author} {\bibinfo {author} {\bibfnamefont {J.~P.~A.}\ \bibnamefont {Novo}}, \bibinfo {author} {\bibfnamefont {P.~V.~P.}\ \bibnamefont {Cunha}},\ and\ \bibinfo {author} {\bibfnamefont {C.~A.~R.}\ \bibnamefont {Herdeiro}},\ }\bibfield  {title} {\bibinfo {title} {{Hypershadows of higher dimensional black objects: a case study of cohomogeneity-one d=5 Myers-Perry}},\ }\Eprint {https://arxiv.org/abs/2410.05390} {arXiv:2410.05390 [gr-qc]}  (\bibinfo {year} {2024})\BibitemShut {NoStop}%
\bibitem [{\citenamefont {Myers}\ and\ \citenamefont {Perry}(1986)}]{Myers:1986un}%
  \BibitemOpen
  \bibfield  {author} {\bibinfo {author} {\bibfnamefont {R.~C.}\ \bibnamefont {Myers}}\ and\ \bibinfo {author} {\bibfnamefont {M.~J.}\ \bibnamefont {Perry}},\ }\bibfield  {title} {\bibinfo {title} {{Black Holes in Higher Dimensional Space-Times}},\ }\href {https://doi.org/10.1016/0003-4916(86)90186-7} {\bibfield  {journal} {\bibinfo  {journal} {Annals Phys.}\ }\textbf {\bibinfo {volume} {172}},\ \bibinfo {pages} {304} (\bibinfo {year} {1986})}\BibitemShut {NoStop}%
\bibitem [{\citenamefont {Myers}(2012)}]{Myers:2011yc}%
  \BibitemOpen
  \bibfield  {author} {\bibinfo {author} {\bibfnamefont {R.~C.}\ \bibnamefont {Myers}},\ }\bibinfo {title} {{Myers\textendash{}Perry black holes}},\ in\ \href@noop {} {\emph {\bibinfo {booktitle} {{Black holes in higher dimensions}}}},\ \bibinfo {editor} {edited by\ \bibinfo {editor} {\bibfnamefont {G.~T.}\ \bibnamefont {Horowitz}}}\ (\bibinfo {year} {2012})\ pp.\ \bibinfo {pages} {101--133},\ \Eprint {https://arxiv.org/abs/1111.1903} {arXiv:1111.1903 [gr-qc]} \BibitemShut {NoStop}%
\bibitem [{\citenamefont {Vasudevan}\ \emph {et~al.}(2005)\citenamefont {Vasudevan}, \citenamefont {Stevens},\ and\ \citenamefont {Page}}]{Vasudevan:2004mr}%
  \BibitemOpen
  \bibfield  {author} {\bibinfo {author} {\bibfnamefont {M.}~\bibnamefont {Vasudevan}}, \bibinfo {author} {\bibfnamefont {K.~A.}\ \bibnamefont {Stevens}},\ and\ \bibinfo {author} {\bibfnamefont {D.~N.}\ \bibnamefont {Page}},\ }\bibfield  {title} {\bibinfo {title} {{Particle motion and scalar field propagation in Myers-Perry black hole spacetimes in all dimensions}},\ }\href {https://doi.org/10.1088/0264-9381/22/7/017} {\bibfield  {journal} {\bibinfo  {journal} {Class. Quant. Grav.}\ }\textbf {\bibinfo {volume} {22}},\ \bibinfo {pages} {1469} (\bibinfo {year} {2005})},\ \Eprint {https://arxiv.org/abs/gr-qc/0407030} {arXiv:gr-qc/0407030} \BibitemShut {NoStop}%
\bibitem [{\citenamefont {Carter}(1968)}]{Carter:1968rr}%
  \BibitemOpen
  \bibfield  {author} {\bibinfo {author} {\bibfnamefont {B.}~\bibnamefont {Carter}},\ }\bibfield  {title} {\bibinfo {title} {{Global structure of the Kerr family of gravitational fields}},\ }\href {https://doi.org/10.1103/PhysRev.174.1559} {\bibfield  {journal} {\bibinfo  {journal} {Phys. Rev.}\ }\textbf {\bibinfo {volume} {174}},\ \bibinfo {pages} {1559} (\bibinfo {year} {1968})}\BibitemShut {NoStop}%
\bibitem [{\citenamefont {Frolov}\ and\ \citenamefont {Kubiznak}(2007)}]{Frolov:2006dqt}%
  \BibitemOpen
  \bibfield  {author} {\bibinfo {author} {\bibfnamefont {V.~P.}\ \bibnamefont {Frolov}}\ and\ \bibinfo {author} {\bibfnamefont {D.}~\bibnamefont {Kubiznak}},\ }\bibfield  {title} {\bibinfo {title} {{Hidden Symmetries of Higher Dimensional Rotating Black Holes}},\ }\href {https://doi.org/10.1103/PhysRevLett.98.011101} {\bibfield  {journal} {\bibinfo  {journal} {Phys. Rev. Lett.}\ }\textbf {\bibinfo {volume} {98}},\ \bibinfo {pages} {011101} (\bibinfo {year} {2007})},\ \Eprint {https://arxiv.org/abs/gr-qc/0605058} {arXiv:gr-qc/0605058} \BibitemShut {NoStop}%
\bibitem [{\citenamefont {Diemer}\ \emph {et~al.}(2014)\citenamefont {Diemer}, \citenamefont {Kunz}, \citenamefont {L\"ammerzahl},\ and\ \citenamefont {Reimers}}]{Diemer:2014lba}%
  \BibitemOpen
  \bibfield  {author} {\bibinfo {author} {\bibfnamefont {V.}~\bibnamefont {Diemer}}, \bibinfo {author} {\bibfnamefont {J.}~\bibnamefont {Kunz}}, \bibinfo {author} {\bibfnamefont {C.}~\bibnamefont {L\"ammerzahl}},\ and\ \bibinfo {author} {\bibfnamefont {S.}~\bibnamefont {Reimers}},\ }\bibfield  {title} {\bibinfo {title} {{Dynamics of test particles in the general five-dimensional Myers-Perry spacetime}},\ }\href {https://doi.org/10.1103/PhysRevD.89.124026} {\bibfield  {journal} {\bibinfo  {journal} {Phys. Rev. D}\ }\textbf {\bibinfo {volume} {89}},\ \bibinfo {pages} {124026} (\bibinfo {year} {2014})},\ \Eprint {https://arxiv.org/abs/1404.3865} {arXiv:1404.3865 [gr-qc]} \BibitemShut {NoStop}%
\bibitem [{\citenamefont {Chandrasekhar}(1985)}]{Chandrasekhar:1985kt}%
  \BibitemOpen
  \bibfield  {author} {\bibinfo {author} {\bibfnamefont {S.}~\bibnamefont {Chandrasekhar}},\ }\href@noop {} {\emph {\bibinfo {title} {{The mathematical theory of black holes}}}}\ (\bibinfo {year} {1985})\BibitemShut {NoStop}%
\bibitem [{\citenamefont {Anderson}\ \emph {et~al.}(1998)\citenamefont {Anderson}, \citenamefont {Jackson},\ and\ \citenamefont {Sitharam}}]{Anderson1998447}%
  \BibitemOpen
  \bibfield  {author} {\bibinfo {author} {\bibfnamefont {B.}~\bibnamefont {Anderson}}, \bibinfo {author} {\bibfnamefont {J.}~\bibnamefont {Jackson}},\ and\ \bibinfo {author} {\bibfnamefont {M.}~\bibnamefont {Sitharam}},\ }\bibfield  {title} {\bibinfo {title} {{Descartes' rule of signs revisited}},\ }\href {https://doi.org/10.2307/3109807} {\bibfield  {journal} {\bibinfo  {journal} {American Mathematical Monthly}\ }\textbf {\bibinfo {volume} {105}},\ \bibinfo {pages} {447} (\bibinfo {year} {1998})}\BibitemShut {NoStop}%
\bibitem [{Note1()}]{Note1}%
  \BibitemOpen
  \bibinfo {note} {Note that a similar behaviour also occurs in the extremal Kerr spacetime where the radial location of the inner light ring coincides with that of the horizon, although they are at some finite proper distance.}\BibitemShut {Stop}%
\bibitem [{\citenamefont {Cunha}\ \emph {et~al.}(2024)\citenamefont {Cunha}, \citenamefont {Herdeiro},\ and\ \citenamefont {Novo}}]{Cunha:2024ajc}%
  \BibitemOpen
  \bibfield  {author} {\bibinfo {author} {\bibfnamefont {P.~V.~P.}\ \bibnamefont {Cunha}}, \bibinfo {author} {\bibfnamefont {C.~A.~R.}\ \bibnamefont {Herdeiro}},\ and\ \bibinfo {author} {\bibfnamefont {J.~a. P.~A.}\ \bibnamefont {Novo}},\ }\bibfield  {title} {\bibinfo {title} {{Light rings on stationary axisymmetric spacetimes: Blind to the topology and able to coexist}},\ }\href {https://doi.org/10.1103/PhysRevD.109.064050} {\bibfield  {journal} {\bibinfo  {journal} {Phys. Rev. D}\ }\textbf {\bibinfo {volume} {109}},\ \bibinfo {pages} {064050} (\bibinfo {year} {2024})},\ \Eprint {https://arxiv.org/abs/2401.05495} {arXiv:2401.05495 [gr-qc]} \BibitemShut {NoStop}%
\bibitem [{\citenamefont {Hawking}(1972)}]{Hawking:1971vc}%
  \BibitemOpen
  \bibfield  {author} {\bibinfo {author} {\bibfnamefont {S.~W.}\ \bibnamefont {Hawking}},\ }\bibfield  {title} {\bibinfo {title} {{Black holes in general relativity}},\ }\href {https://doi.org/10.1007/BF01877517} {\bibfield  {journal} {\bibinfo  {journal} {Commun. Math. Phys.}\ }\textbf {\bibinfo {volume} {25}},\ \bibinfo {pages} {152} (\bibinfo {year} {1972})}\BibitemShut {NoStop}%
\bibitem [{\citenamefont {Hawking}\ and\ \citenamefont {Ellis}(2023)}]{Hawking:1973uf}%
  \BibitemOpen
  \bibfield  {author} {\bibinfo {author} {\bibfnamefont {S.~W.}\ \bibnamefont {Hawking}}\ and\ \bibinfo {author} {\bibfnamefont {G.~F.~R.}\ \bibnamefont {Ellis}},\ }\href {https://doi.org/10.1017/9781009253161} {\emph {\bibinfo {title} {{The Large Scale Structure of Space-Time}}}},\ Cambridge Monographs on Mathematical Physics\ (\bibinfo  {publisher} {Cambridge University Press},\ \bibinfo {year} {2023})\BibitemShut {NoStop}%
\bibitem [{\citenamefont {Aliev}(2006)}]{Aliev:2006yk}%
  \BibitemOpen
  \bibfield  {author} {\bibinfo {author} {\bibfnamefont {A.~N.}\ \bibnamefont {Aliev}},\ }\bibfield  {title} {\bibinfo {title} {{Rotating black holes in higher dimensional Einstein-Maxwell gravity}},\ }\href {https://doi.org/10.1103/PhysRevD.74.024011} {\bibfield  {journal} {\bibinfo  {journal} {Phys. Rev. D}\ }\textbf {\bibinfo {volume} {74}},\ \bibinfo {pages} {024011} (\bibinfo {year} {2006})},\ \Eprint {https://arxiv.org/abs/hep-th/0604207} {arXiv:hep-th/0604207} \BibitemShut {NoStop}%
\end{thebibliography}%

\end{document}